# Quantum Power Electronics: From Theory to Implementation

**Meysam Gheisarnejad [1,*] and Mohammad-Hassan Khooban [1]**

1 Department of Electrical and Computer Engineering, Aarhus University, Denmark; mhkhoban@gmail.com
* Correspondence: Correspondence: me.gheisarnejad@gmail.com

**Abstract:** While impressive progress has been already achieved in wide-bandgap (WBG) semiconductors such as 4H-SiC and GaN technologies, the lack of intelligent methodologies to control the gate drives prevented to the exploit the maximum potential of semiconductor chips, from obtaining the desired devices operations. Thus, a potent ongoing trend is to design a fast gate driver switching scheme to upgrade the performance of electronic equipment at the system level. To address this issue, this work proposes a novel intelligent scheme for the control of gate driver switching using the concept of quantum computation in machine learning. In particular, the quantum principle is incorporated into deep reinforcement learning (DRL) to address the hardware limitations of conventional computers and the growing amount of data sets. Taking potential benefits of quantum theory, the DRL algorithm influenced by quantum specifications (referred to as QDRL) will not only ameliorate the performance of the native algorithm on traditional computers, but also enhance the progress of relevant research fields like quantum computing and machine learning. To test the practicability and usefulness of QDRL, a dc/dc parallel boost converters feeding constant power loads (CPLs) is chosen as the case study, and several Power Hard-ware-in-the-loop (PHiL) experiments and comparative analysis are given.

**Keywords:** Index Terms— Quantum deep reinforcement learning; wide-bandgap semiconductor; Deep Reinforcement Learning; Power Hard-ware-in-the-loop (PHiL)

## 1. Introduction and Preliminaries

HE CONTINUAL requests for more packed PCB footprints has pushed the power electronic researchers to develop novel configurations with fewer components, compacting technologies, and utilize advanced semiconductors which can promote power density. When providing regulated power to on-board semiconductors, the significance of switching frequency appears immediately clear to systems designers. Accordingly, it is desirable to enhance the regulatory frequency, as this can decrease the level of energy in passive elements (e.g., inductors, resistors, etc.) which can result in a smaller size. The primary transistors were manufactured by Silicon (Si) for many decades although Si suffers significant constraints in relation with the operation temperature, light transmission, switching frequency, etc. [1]. Currently, the maximum breakdown voltage strength of a commercial Si-insulated gate bipolar transistor (IGBT) is 8.4 kV with a limited switching operation, while the permissible temperature of any Si-based device is less than 200 °C [1, 2]. To achieve more superior performance in the terms of efficiency, switching speed, and compactness, wide-bandgap (WBG) semiconductors were developed as a good option to displace conventional Si-based products into electronics equipment.

Among the most popular WBG materials, the silicon carbide (4H-SiC) and gallium nitride (GaN) offer a higher level of breakdown capability, greater heat conductivity, greater switching frequency, and greater carrier saturation drift speed [3]. The first generation of GaN semiconductors, especially high electron mobility transistors (HEMTs), were fabricated in a lateral due to initial challenges in the GaN substrates. Despite the success of radio-frequency (RF) power field [4], the limitations of the peak electric field in the lateral configuration led to the low-voltage band (<650 V) in the GaN-based HEMTs. With the recent progress in the commercial GaN bulk substrates, a higher breakdown voltage $(V_B)$ with thinner drift layers is reachable using the new generations of GaN devices in the vertical structure [5]. The utilization of the two-dimensional electron gas (2DEG) formation has enabled the possibility of high-frequency switching capability in the vertical AlGaN/GaN heterostructures, where mobility values are usually more than 1000 square centimeters $cm^2.V^{-1}.s^{-1}$. Unlike the GaN counterparts, the SiC semiconductors have been adopted as a promising material for low-frequency and high-voltage power applications [6]. Due to their high gate charge and need for high-power gate circuits, SiC devices are often adopted at low frequencies. However, the increasing demand for SiC components faces the bulkiness of the gate driver circuits and design complexity. Ref. [7] demonstrates a 2-kW Class $\Phi_2$ inverter with the SiC MOSFET has a lower volume than the Si-based gate driver. The deployment of integrated SiC-based gate drivers, according to recent state-of-the-art semiconductor studies [8, 9], led to lighter and more compact power electronic devices. While semiconductor chips delivered remarkable progress at the device level, no matching advances have been observed at the system level (drivers, control algorithms, etc.), and as a result, a large portion of that potential is being lost

Proper regulation of the gate driver to accomplish minimization, high efficiency, and size reduction—all of which are related to the switching frequency—is the essential strategy for maximizing the benefits of semiconductors. For the design of intelligent gate drivers, many practical experiments include widely soft-switching techniques to reach enhanced efficiency and high-speed switching using reducing losses which can be derived by zero-voltage or zero-current transitions. To make further advancements without efficiency deterioration, numerous auxiliary-circuit-based (i.e., hardware-based) approaches, like quasi-resonant (QR), multiple-resonant (MR), and series/parallel/series-parallel resonant (SR/PR/SPR) have been introduced [10]. Despite the high efficiency of deploying such auxiliary circuits, their design not only leads to the increasing cost and overall dimension of a power device but also creates conduction loss because of the circulating currents in the auxiliary components. To address these issues, many nonauxiliary-circuit-based methodologies (i.e., software-based) such as single/dual phase-shift (SPS/DPS), phase-shift-modulation (PSM), and trapezoidal-modulation (TZM), which add a higher level of reliability with reduced cost because no unnecessary auxiliary components, were reported to control gate drivers [11]. However, these techniques are complex to regulate, and the zero-voltage-switching (ZVS) may be malfunctioned within the full-power range because of the limited gain ratio, hence their application is limited at low and medium power systems.

With the progress of big data and artificial intelligence, machine learning (ML) has the potential to revolutionize the next generations of power electronic interfaces and make a substantial impact on terms efficiency, .Quantum ML (QML) is a field that aims to fully integrate conventional ML and quantum information processing (QIP) to address and overcome issues that have been seen in algorithmic tasks of conventional ML (e.g., time-consuming, data acquisition, and kernel estimation). On the bias of the algorithmic process, many proposals were made in the context of QML including supervised, unsupervised, and semi-supervised learning. In contrast, fewer state-of-the-art studies pay attention to the development of reinforcement learning (RL) in the QML community, and particularly proposals for quantum process deep RL (QPDRL) are becoming an emerging field. The concept of QPDRL first emerged in 2008 [12] from incorporating the principle of quantum parallelism into traditional reinforcement learning, which provided the right balance between exploration and exploitation and accelerated training as well. Dunjko *et al.* theoretically demonstrated that with the application of QPDRL, a [13] quadratic amelioration in the training efficiency and exponential amelioration in operation will be provided for a wide class of training problems. The QPDRL with multiqubit has also been tested on state-of-the-art superconducting circuits [14] and expanded to include multi-level systems and open quantum dynamics, among other scenarios [15]. Additionally, recent studies [16] have shown that employing quantum Boltzmann machines for RL is advantageous compared to using a conventional one.

Unlike the circuit designs and configurations that have less importance to take the full benefits of GaN and 4H-SiC, the proper control of the gate driver is the key option for benefiting the maximum potentials of advanced semiconductors at the system level. The paradigm shift from today's gate driver control methodologies to innovative ultra-super-fast schemes and design tools is required to offer a higher level of efficiency for power electronic interfaces and motor drivers, resulting in realistic improving compactness and reducing energy consumption.

The primary objective of this work is to design a quantum gate driver with very high-speed motor drivers with control techniques drawing on modern DSP and an understanding of power transistor device physics. Taking benefits of quantum theory, an intelligent gate driver based on Quantum DRL will be designed based on the nature of subatomic particles to conduct the computations several times faster than the most sophisticated algorithms. The real-time examinations based on OPAL-RT setup are accomplished with a comparison to model predictive control.

The remainder of this work is organized as follows. The dynamic modeling of the power electronic test system is presented in Section II. The theory of ultra-local model control is introduced in Section III. In Section IV, the framework of QPDRL is introduced and quantum representations and quantum operations are presented, followed by the algorithm description of QPDRL with specific implementation details. In Section V, experimental results are presented to verify the applicability of the proposed QPDRL algorithm.

## 2. Dynamics Of dc MG with Parallel Boost Converter

Fig. 1 represents the simplified model and typical configuration of a DC power electronic system comprising two boost converters. In the structure, the boost converters are connected as parallel to transfer the power from the DC voltage source (e.g., Photovoltaic, Fuel Cell, etc.) which is connected to the main bus to supply the CPL [17].

In Fig. 1, $L_{b1,b2}$, $C_{b1,b2}$, and $i_{L,b1,b2}$ denote the inductance and capacitor of converters, respectively; $L_{line,1,2}$ and $R_{line,1,2}$ denote the line resistance and line inductance, respectively. $P_{cpl,1,2}$ denotes the CPL's power; and $C_{cpl,1,2}$ denotes the CPL's capacitor. $i_{line1,line2}$ denotes the transmission line current of the converter; $E_1$ and $E_2$ denote the input voltages of converter-1 and converter-2, respectively; $v_{out,i}$ and $v_{out,i}$ are the output signals of the converter-1 and converter-2, respectively.

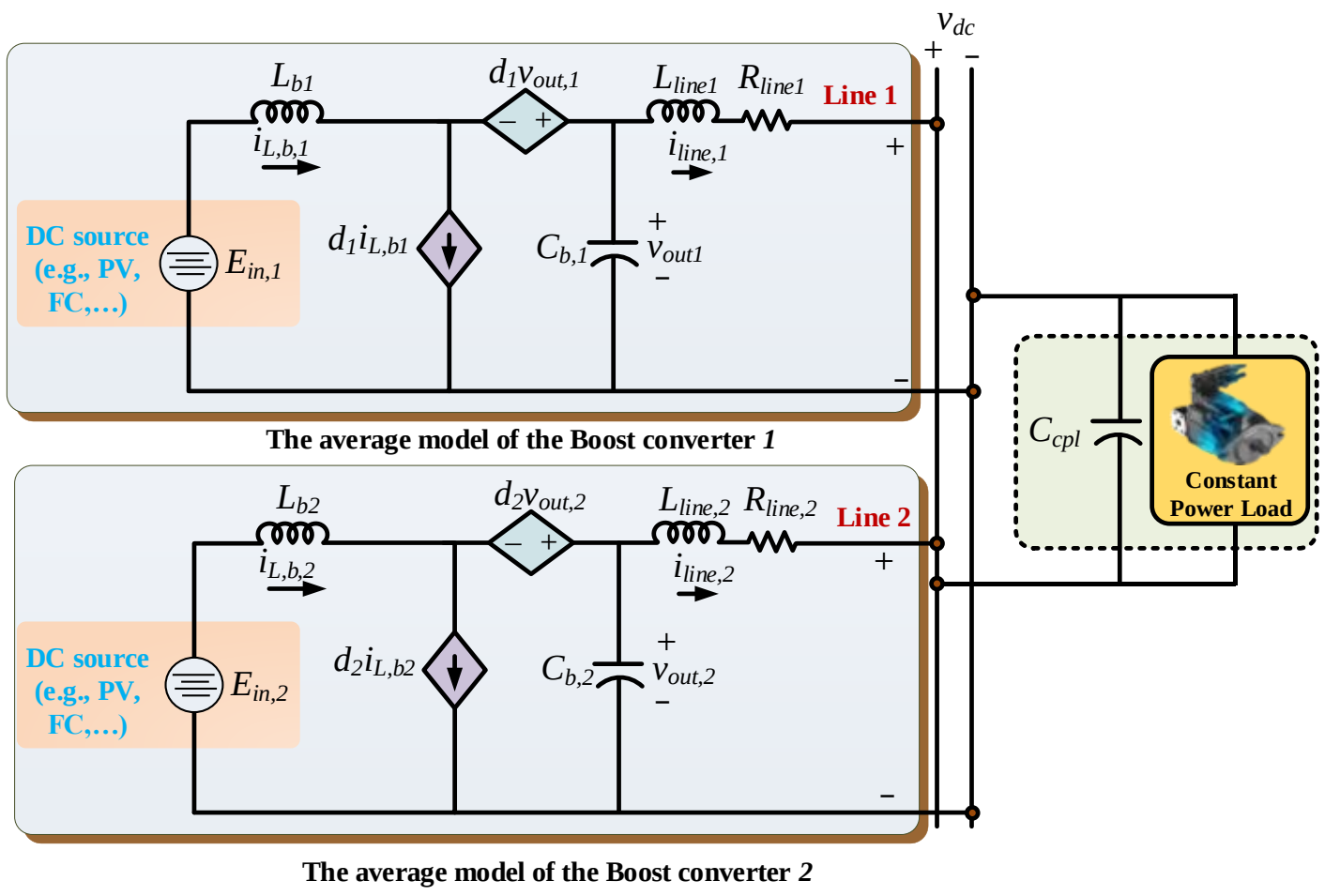

**Figure 1.** Structure of DC power electronic system with two parallel boost converters.

The average model of the converter is expressed as:

$$L_{b,i}\frac{di_{L,b,i}}{dt} = v_{in,i} - (1-d_i)v_{out,i} \tag{1}$$

$$C_{b,i}\frac{dv_{out,i}}{dt} = (1-d_i)i_{L,b,i} - i_{line,i} \tag{2}$$

The equations corresponding to the power converter are given as:

$$L_{line,i}\frac{di_{line,i}}{dt} = v_{out,i} - R_{line,i}i_{line,i} - v_{dc} \tag{3}$$

$$\left(C_{cpl,1} + \cdots + C_{cpl,i}\right)\frac{dv_{dc}}{dt} = \sum_{i=1}^{n} i_{line,i} - \frac{1}{v_{dc}}\sum_{i=1}^{n} P_{cpl,i} \tag{4}$$

The main role of the controller in the power electronic system is to the output voltage $v_{out}$ tracks the reference's voltage $v_{ref.}$.

**3. Structure of Ultra-local model control**

Assume the dynamic model of a system is unknown and can be described by a single-input single-output (SI-SO) system. The ultra-local model (ULM) of the system can be defined to determine the unknown mathematical model. For a SISO system, the input-output $(u, y)$ behavior of the system can be expressed by [18]:

$$\boldsymbol{E}\left(t, y, y^{[1]}, \dots, y^{[\iota]}, u, u^{[1]}, \dots, u^{[\kappa]}\right) = 0 \tag{5}$$

where $\boldsymbol{E}$ is an unknown function with a smooth function of its arguments, $[\iota]$ and $[\kappa]$ are the orders of $u$ and $y$, respectively, and $y$ is defined as:

$$y^{[\vartheta]} = \boldsymbol{E}\left(t, y, y^{[1]}, \dots, y^{[\vartheta-1]}, y^{[\vartheta+1]}, y^{[\iota]}, u, u^{[1]}, \dots, u^{[\kappa]}\right) \tag{6}$$

where $0 < \vartheta \leq \iota$, $\frac{\partial \boldsymbol{E}}{\partial y^{[\vartheta]}} \neq 0$. A numerical model of (6) can be described by ULM for a very short lapse of the interval, given as:

$$y^{[\vartheta]}(t) = \chi u(t) + \mathcal{F} \tag{7}$$

where $\chi \in \mathbb{R}$ is a non-physical factor selected such that the $y^{[\vartheta]}(t)$ and $\chi u(t)$ will have the same order, term $\mathcal{F}$ includes all the unmodelled dynamics which are often calculated by the algebraic approach.

The expression of the law control for a unit derivative order can be formed as an intelligent proportional-integral (iPI) controller:

$$u(t) = \frac{1}{\chi}\left(-\hat{\mathcal{F}} + \dot{y}^{*}(t) + K_p e(t) + K_i \int e(t)\, dt\right) \tag{8}$$

where $\hat{\mathcal{F}}$ denotes the estimation of unmodeled dynamics, $K_p$ and $K_i$ are the tunable parameters of the iPI controller.

**4. Quantum Deep Reinforcement Learning**

The reduction of reinforcement learning (RL) actions is necessary to counteract the dimensionality curse, however fewer command numbers may result in incorrect control signals. By adopting the Quantum mechanism, a higher level of control signals can be generated than the action sets in the standard RL. In this way, deep belief networks (DBNs) can be adopted to predict the next systemic information which will lead to accurate control commands.

### 4.1. Principal of quantum theory

The quantum bit (qubit), which functions similarly to traditional bits, is regarded as the core idea of quantum computation. In quantum theory, two basic levels of the qubit are defined by $|0\rangle$ and $|1\rangle$, which are equivalent to the traditional bit states of 0 and 1. However, a qubit $|\psi\rangle$ can be expressed based on a superposition state of $|0\rangle$ and $|1\rangle$ regarding complex coefficients of $\alpha$ and $\beta$, given as [19]:

$$|\psi\rangle = \alpha|0\rangle + \beta|1\rangle \tag{9}$$

An RL agent interact with its environment which is mathematically formulated by Markov Decision Process (MDP). The $i^{\text{th}}$ output of the quantum process based on DRL can be obtained by the following descriptions:

$$\delta'_{i,out} = \delta_k + \frac{a_{(k+1)} + a_{(k-1)}}{2}\left(q_i(a) - \frac{1}{2}\right) \tag{10}$$

where $a_{(k+1)}$ and $a_{(k-1)}$ denote the actions of set $A$; the selected action among the possible actions of $A$ is shown by $a_k$. Likewise, $0 \le q_i(a) \le 1$ is the output probability which is expressed by:

$$q_i(\delta) = |\delta_\delta^{N_A}\rangle = \sum_{\delta=00\ldots0}^{\overbrace{11\ldots1}^{N_Q}} C_\delta|\delta\rangle \tag{11}$$

where $\sum_{\delta=00\ldots0}^{\overbrace{11\ldots1}^{N_Q}}|C_\delta|^2 = 1$; $C_\delta$ is a complex factor; $|C_a|^2$ denotes occurrence probability of $|\delta\rangle$ in the action $|\delta_\delta^{N_A}\rangle$; $N_Q$ is the quantum bit count.

### 4.2. Principal of RL

Each subordinate RL is made from four parts: i) a Q-value renovated activity, ii) a P-value renovated activity, iii) an action selection mechanism, and iv) a quantum procedure.

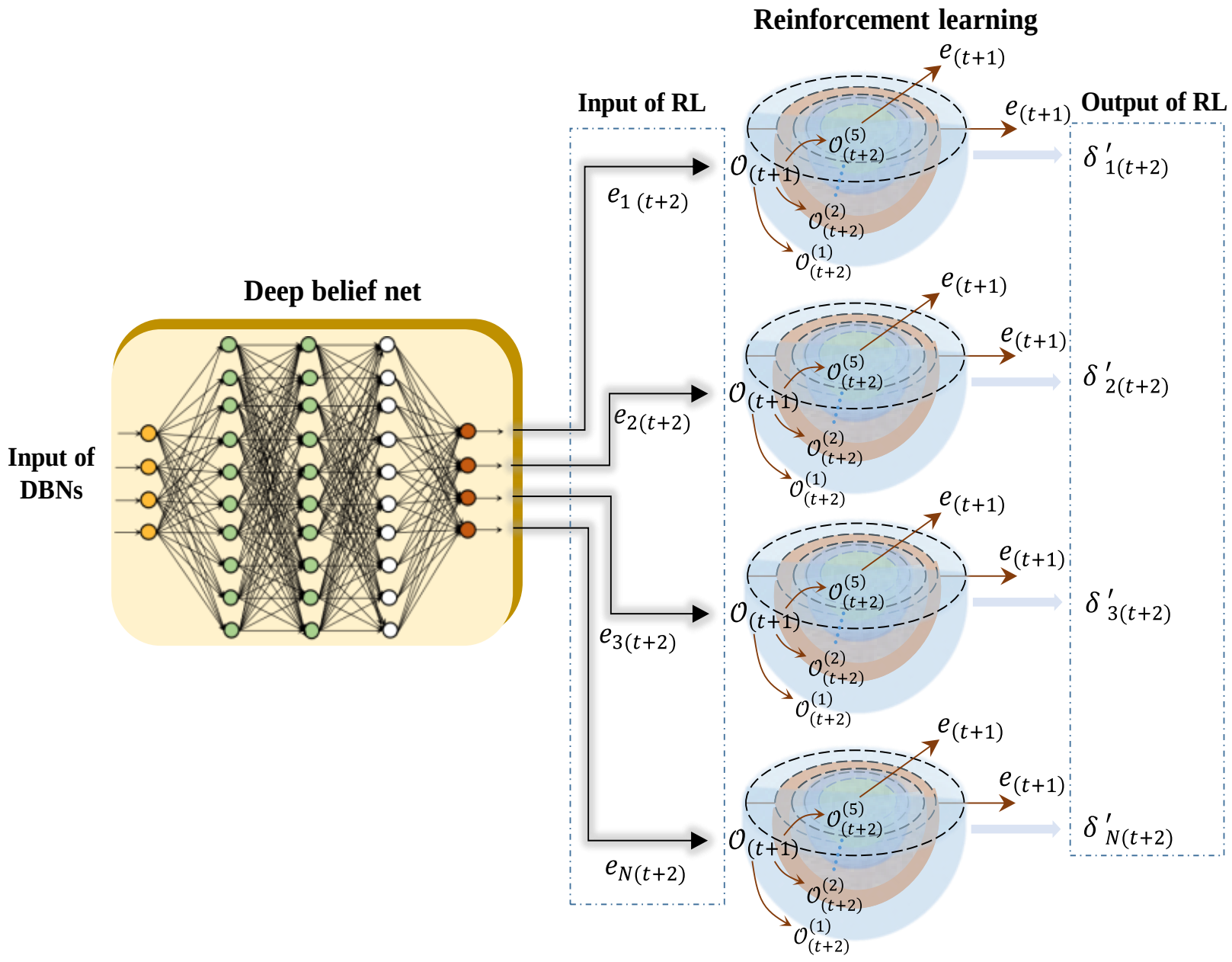

Fig. 2. Training process of quantum process deep RL with deep belief net.

With the state prediction of DBNs, the Q-value of each subsidiary component can be updated. In the next stage, the P-value will be updated by P-value process, providing P-value mechanism to acquire the action index. Then, the actions will be chosen from the actions set $A$. In the final stage of QDRL, the quantum process generates the real actions. The control actions of QDRL-agent are provided by updating the Q-value matrix and $P$-value matrix, given as [16]:

$$Q_{RL}(\mathcal{O}',\delta) = Q_{RL}(\mathcal{O},\delta) + \zeta_{RL}\left(P_{RL}(\mathcal{O},\mathcal{O}',\delta) + \gamma_{RL}\max_{\delta\in A} Q_{RL}(\mathcal{O},\delta') - Q_{RL}(\mathcal{O},\delta)\right) \tag{12}$$

$$P_{RL}(\mathcal{O}',\delta) = \begin{cases} P_{RL}(\mathcal{O},\delta) - \mu_{RL}\big(1 - P_{RL}(\mathcal{O},\mathcal{O}',\delta)\big), if\ \delta' = \delta \\ P_{RL}(\mathcal{O},\delta) - (1-\mu_{RL}), if\ \delta' \neq \delta \end{cases} \tag{13}$$

In the above equations, $\zeta_{RL}$, $\gamma_{RL}$, and $\mu_{RL}$ denote the learning factor, discount factor, and updated factor, respectively. Generally, all the parameters of $\zeta_{RL}$, $\gamma_{RL}$, and $\mu_{RL}$ are configured in the band of $\zeta_{RL}$, $\gamma_{RL}$, and $\mu_{RL} \in [0,1]$. The terms of $\mathcal{O}$, $\delta$, and $\mathcal{O}'$ are the current state, action, and anticipated next state, correspondingly. The reward function of QPDRL for a wide range of power electronic test systems can be defined as:

$$R_{RL}(\mathcal{O},\mathcal{O}',\delta) = \begin{cases} \dfrac{c_2}{|v_o(t) - v_{ref.}(t)|} & if\ \left(v_o(t) - V_{Ref}(t)\right) < 0.05 \\ -c_1|v_o(t) - V_{Ref}(t)| & if\ \left(v_o(t) - V_{Ref}(t)\right) > 0.05 \end{cases} \tag{14}$$

where $c_1$ and $c_2$ denote the constant coefficients.

*4.3. Deep belief nets (BBNs) based on Restricted Boltzmann Machines*

The deep belief nets (DBNs) are a potent generative model which employs a deep structure of multiple restricted Boltzmann machines (RBM). In the DBNs, it is assumed that the quantity of hidden layers $N_{Layer}$ and hidden units $N_{Hidden}$ be equal. By considering the set of $\boldsymbol{\theta} = \{\boldsymbol{W}^R, \boldsymbol{b}_v, \boldsymbol{b}_h\}$, one can have [20]:

$$E(\boldsymbol{v},\boldsymbol{h};\boldsymbol{\theta}) = -\sum_{i=1}^{N_{Layer}} b_{vi}v_i - \sum_{j=1}^{N_{Hidden}} b_{hi}h_i - \sum_{i=1}^{N_{Layer}}\sum_{j=1}^{N_{Hidden}} v_i w_{ij}^R h_j \tag{15}$$

where $W^R$ represents a linking matrix of RBM; $b_v$ represents the biases of hidden neurons; $b_h$ represents the visible neurons; the vectors of $\boldsymbol{v} = (v_1, v_2, \dots, v_{N_{Layer}})$ and $\boldsymbol{h} = (h_1, h_2, \dots, h_{N_{Hidden}})$ represent the property of visible and hidden layers.

The distribution of marginal probability based on the term $v$ is given as:

$$P(\boldsymbol{v};\theta) = \frac{\sum_h e^{-E(v,h;\theta)}}{\sum_{v,h} e^{-E(v,h;\theta)}} \tag{16}$$

For the hidden and visible units, the dynamic probability distribution can be calculated by employing a Gibbs sampling approach, given as:

$$P\big(h_j = 1|\boldsymbol{v};\boldsymbol{\theta}\big) = \frac{1}{1 + e^{\left(-\left(b_j + \sum_i v_i w_{ij}^R\right)\right)}} \tag{17}$$

$$P\big(v_j = 1|\boldsymbol{h};\boldsymbol{\theta}\big) = \frac{1}{1 + e^{\left(-\left(d_i + \sum_j v_j w_{ij}^R\right)\right)}} \tag{18}$$

A detailed illustration of the QPDRL is depicted in Fig. 2.

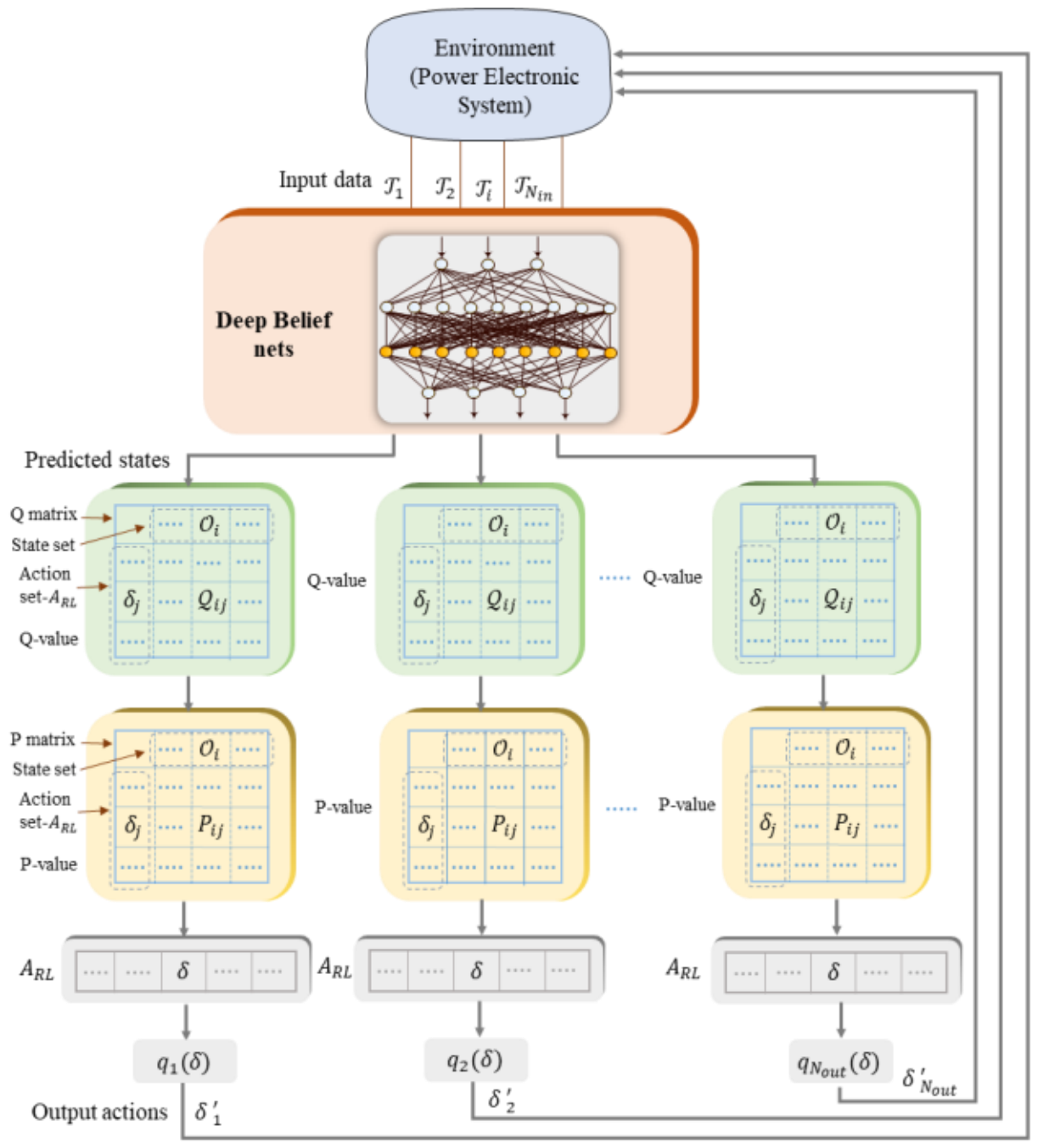

**Figure 3.** Architecture of Quantum process based on deep reinforcement learning.

*4.4. Training of QPDRL*

For the power electronic test system, the terms of $v_{o1}$, $v_{o2,q}$, $\frac{v_{o1}}{dv_{o1}}$ and $\frac{v_{o2}}{dv_{o2}}$ are considered as the input data parameters of the QPDRL. The four variables of output voltages are estimated by the deep nets, i.e., $e_1$, $e_2$, $e_3$, and $e_4$. The training framework of the QPDRL is illustrated in Fig. 3 [16]. By training deep belief nets, the accurate prediction of system states is realized. The control actions generated in the output of QPDRL adjust the control coefficients of the ULM controller, i.e., $K_p$ and $K_i$.

## 5. Real-Time Verifications and Comparisons

The DC power electronic system of this paper is developed in a laboratory OPAL-RT setup to conduct Hardware-in-the-Loop (HiL) simulations for real-time examinations of the proposed control methodology. For comparison, the real-time outcomes of the ULM controller based on QPDRL are compared with the model predictive control scheme.

The values of inductance of the boost converter $L_{b1,b2}$ are set to $6\ mH$; The values of resistance of bus line-1 and bus line-2 are given as $0.8\ \Omega$ and $0.5\ \Omega$, respectively; the values of the capacitor of the boost converter $C_{b1,b2}$ are $200\ \mu F$; inductance of bus line-1 $L_{line,1}$ and line-2 $L_{line,2}$ are as $1.9\ mH$ and $1.2\ mH$, respectively.

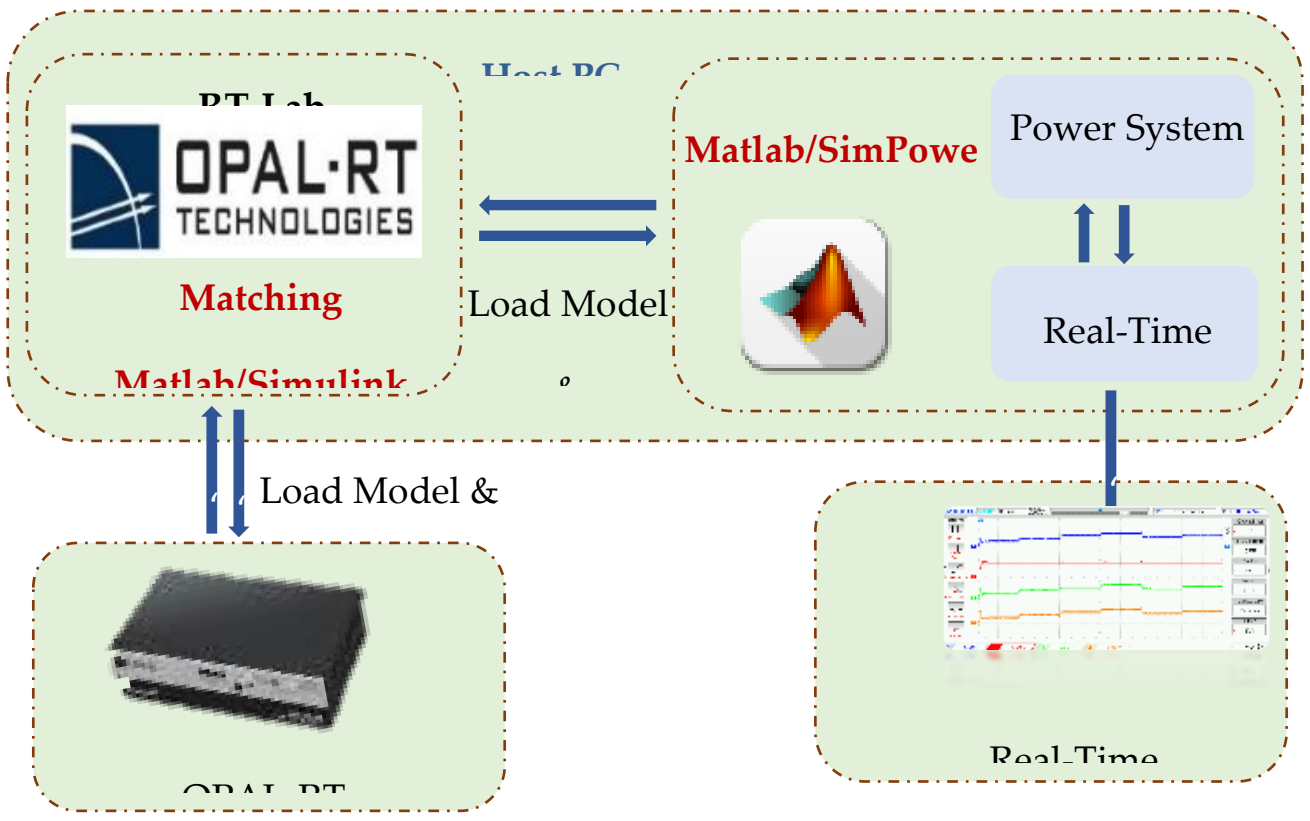

**Figure 4.** Illustration of experimental procedure.

*5.1. Case study* $(i)$ *(under change in CPL's power):*

The DC source voltage for both the converters is set as 500 [V] while the output voltages of CPL will be regulated on $V_{ref.}$ =750 [V]. It should be noted that change in the CPL's power impose the worst condition of instability on the power electronic systems. Thus, a time-varying CPL's power for 0 to 2.4s is applied on the power-electronic converter, given as:

$$P_{cpl} = \begin{cases} 20\ kw & [0s\ 0.4s) \\ 30\ kw & [0.4s\ 0.8s) \\ 40\ kw & [0.8s\ 1.2s) \\ 50\ kw & [1.2s\ 1.6s) \\ 35\ kw & [1.6s\ 2s) \\ 45\ kw & [2s\ 2.4s) \end{cases} \tag{19}$$

The CPL's power, voltage, and current waveforms of HiL setup under CPL's power of (19) for MPC and the proposed controller are shown in Fig. 5(a) and Fig. 5(b), respectively. It is evident that the outcomes of the proposed ULM controller (realized based on QPDRL) experience lower fluctuations than the MPC. Therefore, a lower level of robustness can be reached by the proposed ULM controller-based QPDRL than other state-of-the-art schemes from the stability point of view.

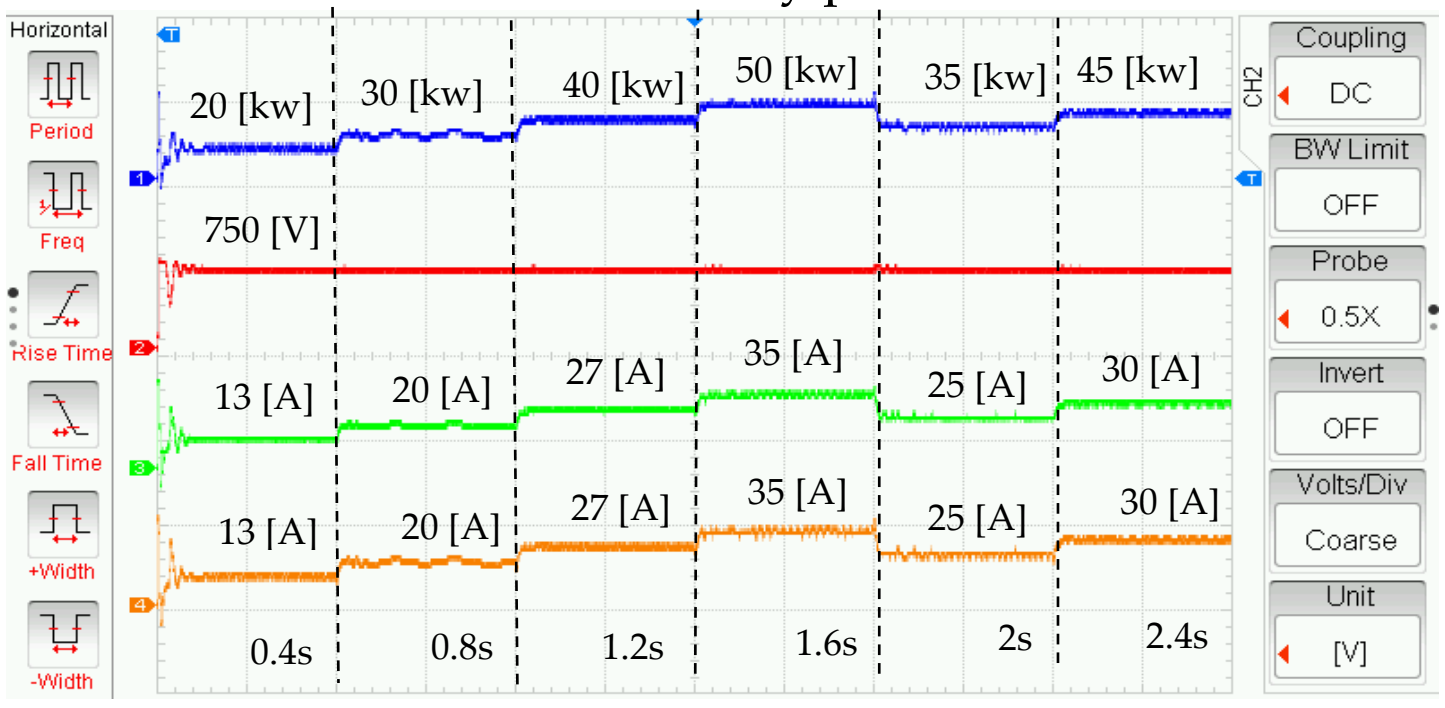

(a)

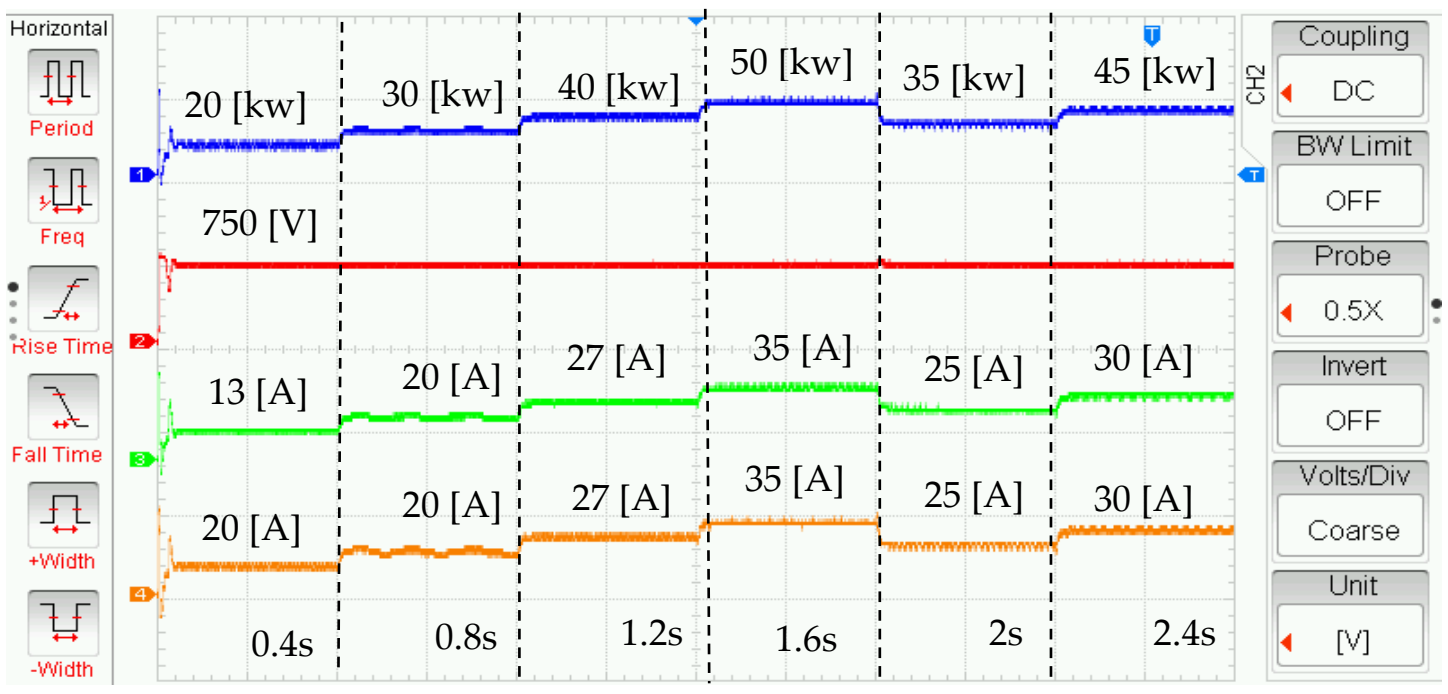

(b)

**Figure 5.** Real-time waveforms of a) MPC, and b) proposed controller under case-study $(i)$: '–' represents CPL's power, '–' represents the voltage, '–' represents current of line 1, and '–' represents current of line 2.

*5.2. Case study* $(ii)$ *(under a change in CPL's power and reference's voltage):*

In this case, the reference voltage and power of CPLs are changed simultaneously during the experiment to examine the performance of the designed controller in more severe conditions. The reference voltage is set to $V_{ref.}$ = 680 [V] for [0 0.8s), $V_{ref.}$ = 800 [V] for [0.8s 1.6s), and $V_{ref.}$ = 600 [V] for [1.6s 2.4s]. The values of CPL's power, voltage, and current outputs of the power electronic test-system with parallel boost converters under changes in the CPL's power and reference voltage for MPC and the proposed controller are depicted in Fig. 6(a) and Fig. 6(b), respectively.

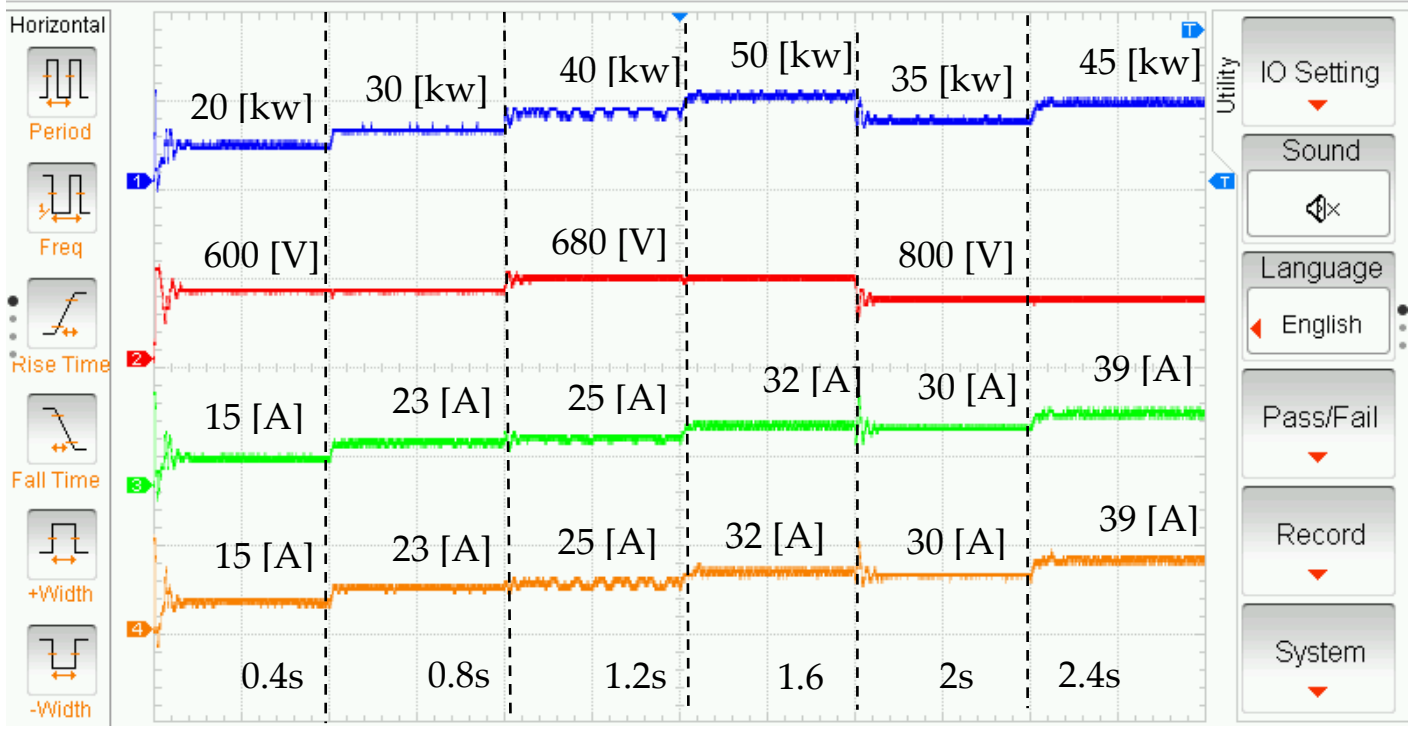

(a)

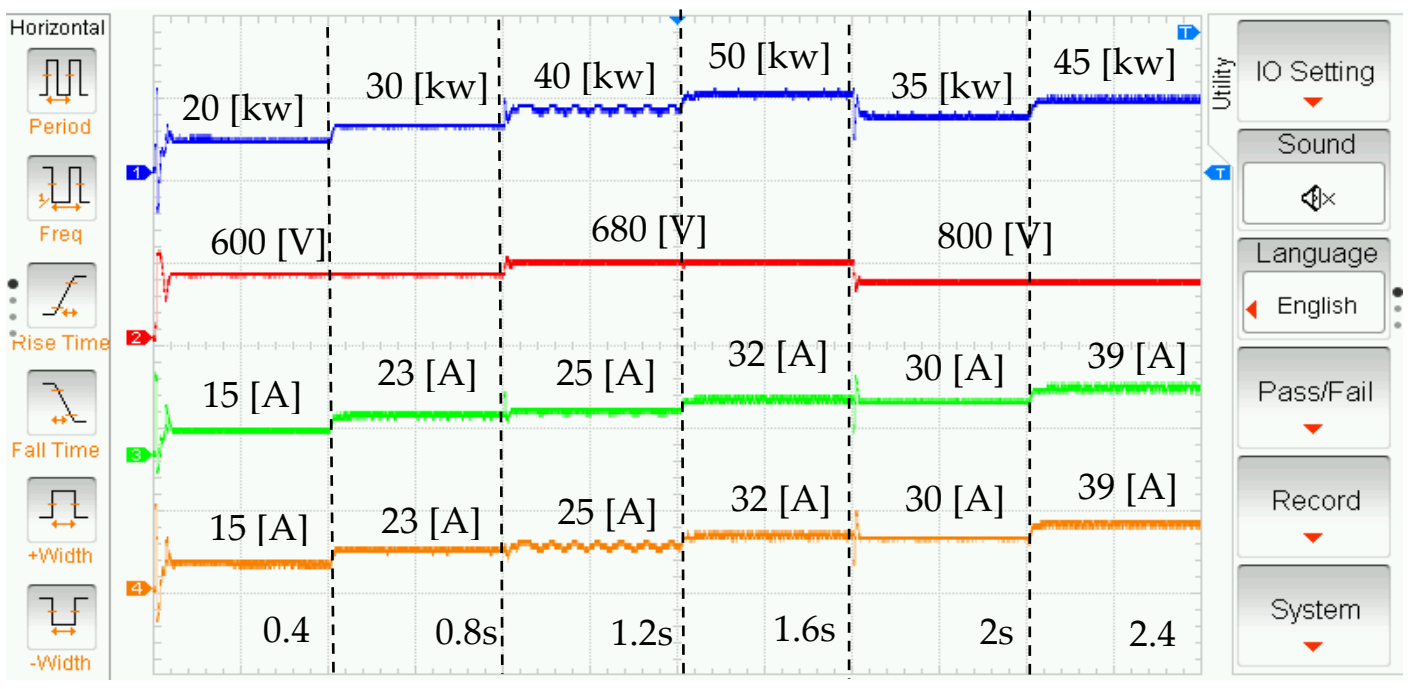

(b)

**Figure 6.** Real-time waveforms of a) MPC, and b) proposed controller under case-study (*ii*): '–' represents CPL's power, '–' represents the voltage, '–' represents current of line 1, and '–' represents current of line 2.

## 6. Conclusion

The suggested quantum process-based-DRL scheme has been exhaustively evaluated by means of HiL experiments accomplished in a 20 W DC multi-converter supplying CPL. To appraise the actual application of QPDRL beyond the proof of concept presented in this work, it is essential to compare QPDRL with many state-of-the-art RL algorithms, in a power electronic test system. The HiL experiments using the OPAL-RT setup reveal that a rich impressibility of the quantum QPDRL contributes to reaching accurate voltage reg-ulation for the parallel boost converters. Despite severe instability imposed by the time-varying CPL, the proposed scheme is still stable while providing a higher level of output regulatio